\documentclass[pra,10pt,a4paper,superscriptaddress,twocolumn,showpacs,notitlepage]{revtex4}
\usepackage{times}
\usepackage{amsmath}
\usepackage{amssymb}
\usepackage{graphicx}
\newcommand{\bb}{\bibitem}
\newcommand{\be}{\begin{equation}}
\newcommand{\ee}{\end{equation}}
\newcommand{\ben}{\begin{eqnarray}}
\newcommand{\een}{\end{eqnarray}}
\newcommand{\bes}{\begin{subequations}}
\newcommand{\ees}{\end{subequations}}

\begin{document}
\title{Modulation of breathers in cigar-shaped Bose-Einstein condensates}
\author{W. B. Cardoso}
\affiliation{Instituto de F\'{\i}sica, Universidade Federal de Goi\'{a}s, 74.001-970, 
Goi\^{a}nia, Goi\'{a}s, Brazil.}
\author{A. T. Avelar}
\affiliation{Instituto de F\'{\i}sica, Universidade Federal de Goi\'{a}s, 74.001-970, 
Goi\^{a}nia, Goi\'{a}s, Brazil.}
\author{D. Bazeia}
\affiliation{Departamento de F\'{\i}sica, Universidade Federal da Para\'{\i}ba,
58.059-900, Jo\~{a}o-Pessoa, Para\'{\i}ba, Brazil.}
\pacs{03.75.-b, 03.75.Lm, 05.45.Yv, 67.85.Hj}

\begin{abstract}
We present new solutions to the nonautonomous nonlinear Schr\"{o}dinger equation
that may be realized through convenient manipulation of Bose-Einstein condensates.
The procedure is based on the modulation of breathers through an analytical 
study of the one-dimensional Gross-Pitaevskii equation, which is known 
to offer a good theoretical model to describe quasi-one-dimensional 
cigar-shaped condensates. Using a specific {\it Ansatz}, we transform 
the nonautonomous nonlinear equation into an autonomous one, which 
engenders composed states corresponding to solutions localized in space, 
with an oscillating behavior in time. Numerical simulations confirm stability
of the breathers against random perturbation on the input profile of the solutions.
\end{abstract}

\maketitle

%%%%%%%%%%%%%%%%%%%%%%%%
%%%%%%%%%%%%%%%%%%%%%%%%
\section{Introduction}

Breathers are solutions of nonlinear equations with a 
profile which is usually localized in space and periodic in time, 
even for equations with constant coefficients. Such solutions usually
come from a compound of two or more solitons located at the same spatial position.
They have been originally introduced in the sine-Gordon 
equation \cite{Ablowitz74}, but they can also be found in other 
scenarios, controlled by the modified Korteweg-de Vries equation 
\cite{Drazin89}, the Davey-Stewartson \cite{Tajiri00} and the 
nonlinear Schr\"{o}dinger equation \cite{K03}. In various physical 
systems, which are well described by nonlinear equations, breathers 
directly affect their electronic, magnetic, optical, vibrational 
and transport properties, such as in Josephson superconducting 
junctions \cite{1,2}, charge density wave systems \cite{3}, 
4-methyl-pyridine crystals \cite{4}, metallic nanoparticles \cite{Meta}, 
conjugated polymers \cite{5}, micromechanical oscillator arrays 
\cite{Micro}, antiferromagnetic Heisenberg chains \cite{6,7}, and 
semiconductor quantum wells \cite{Hass09}.

After the first experimental manipulation of Bose-Einstein condensates (BECs) 
\cite{E1,E2,E3}, studies on bright 
and dark solitons \cite{DS1,DS2,BS1,BS2} have triggered a lot of new investigations,
with a diversity of scenarios being proposed and tested \cite{K03,R1,R2,B2,B3,B5}. 
In particular, the presence of experimental techniques for manipulating the
strength of the effective interaction between trapped atoms \cite{Rob} 
leads us to believe that in BECs we have an excellent opportunity to 
investigate breathers of atomic matter waves taking advantage of 
Feshbach-resonance management \cite{ino,HK,K3,M04,PG06,H07,M05,S04,Y09}. Thus, in this work the 
main aim is to show that breathers can be modulated 
in BECs, presenting a diversity of patterns.
We offer an analytical study of such breather solutions of the Gross-Pitaevskii 
equation (GPE) with both the potential and the parameters that control 
the strength of the nonlinearities being time-dependent. This makes the 
GPE nonautonomous and very hard to solve in general. However, we employ 
an {\it Ansatz} which transforms the nonautonomous GPE into an autonomous 
nonlinear Schr\"odinger equation (NLSE) which is easier to be studied.

Since the one-dimensional GPE is a 
very good theoretical model to describe quasi-one-dimensional cigar-shaped BECs \cite{B2,B3}, 
we then investigate the presence of the breather behavior in BECs under the 
assumption that $a_s\ll a_{\bot}\ll\lambda\ll a_{||}$, for $a_s$ being the 
scattering length, $\lambda$ the spatial scale of the wave packet, and 
$a_{\bot}$ and $a_{||}$ being the characteristics transverse and longitudinal 
trap lengths, respectively. Also, in this work we use potentials and 
nonlinearities which are typical of BECs, and we believe that the results 
obtained below may stimulate new experiments in condensates.

In the recent literature, attention has also been devoted to the case of 
discrete breathers, in arrays of BECs \cite{DB1,DB2,DB3}. However, when 
one looks for continuous dynamics, with the GPE having variable coefficients 
depending on the space and time coordinates, analytical solution is not a 
trivial issue. Here we can point a numerical study  \cite{M05} which emphasizes 
the stability properties of the breather excitations in a three-dimensional 
BEC with Feshbach-resonance management of the scattering length and confined 
by a one-dimensional optical lattice. We believe that such excitations are 
made localized by the addition of an external potential, usually created by an optical 
lattice used to trap them in the lattice. 

In the present paper, however, 
we use a genuine breather solution, i.e., with oscillating behavior in time even 
when the nonlinear equation presents only constant coefficients (i.e., without modulation). 
The profile of these breathers are shown in Fig.~\ref{F1}. We also show that they can be modulated in space 
and time, controlled by the trapping potential and the strength of the cubic nonlinearity, as we show below
for a diversity of situations. The procedure is similar to the former investigations \cite{SH,SH2,SHB,BB,BBC,ABC},
but here we allow for the autonomous NLSE to be partial differential equation, and this introduces an interesting
difference which we explore below.

We start the investigation in the next section, where we present the general formalism, following as close as possible 
the two recent works \cite{BB,ABC}. In Sec.~III we illustrate the formalism with distinct examples, identified by the
trapping potential and cubic nonlinearity, with a diversity of solutions. We also show in Sec. IV, through numerical
simulation, that the solutions found in Sec.~III are all stable under random perturbation of 5\% in their input profile.
There we also comment of the case of nonpolynomial nonlinearity, which can be used to compensate for the one-dimensional
approximation of real condensates. There we show that the one-dimensional approximation used in this work is reliable.
We finish the work in Sec.~V, where we introduce the ending comments.

%%%%%%%%%%%%%%%%%%%%%
\section{Formalism}
%%%%%%%%%%%%%%%%%%%%%
We start with the one-dimensional GPE with the general form
\begin{equation}
i\psi _{t}=-\frac{1}{2}\psi _{xx}+V(x,t)\psi +P(|\psi|^2)\psi ,
\label{GPE}
\end{equation}
where $V(x,t)$ is the potential and $P(|\psi|^2)$ is given by
\begin{equation}
P(|\psi|^2)=\sum_{n=1}^N g_{2n+1}(t)|\psi|^{2n}.
\end{equation}
The coefficients $g_{2n+1}(t)$ of the polynomial function above describe the strength 
of the nonlinearities present in the system. Here we are using
standard notation, in which $x$, $t$ and all other quantities are dimensionless.

Our goal is to find breather solutions of the Eq.(\ref{GPE}). Using the \emph{Ansatz}
\begin{equation}
\psi (x,t)=\rho (t)e^{i\eta (x,t)}\Phi (\zeta (x,t),\tau (t)),
\label{An}
\end{equation}
we can change the nonautonomous GPE for an autonomous NLSE that may engender 
the breather solutions; it is given by
\begin{equation}
i\Phi _{\tau }=-\frac{1}{2}\Phi _{\zeta \zeta } + Q(|\Phi|^2)\Phi ,
\label{NLSE}
\end{equation}
where $Q(|\Phi|^2)$ has the form
\begin{equation}
Q(|\Phi|^2)=\sum_{n=1}^N G_{2n+1}|\Phi|^{2n}.
\end{equation}
The coefficients $G_{2n+1}$ are now constant parameters. Using \eqref{An} in 
\eqref{GPE} leads to \eqref{NLSE}, for $\rho$, $\eta$, $\zeta$ and $\tau$ 
obeying the following equations
\bes
\ben
&&\tau _{t}=\zeta _{x}^{2},\label{a}
\\
&&\eta _{x}=-\frac{\zeta _{t}}{\zeta _{x}},\label{b}
\\
&&\rho _{t}+\frac{\rho \eta _{xx}}{2}=0.\label{c}
\een\ees
Here the potential and nonlinearities assume the form
\begin{equation}
V(x,t)=-\eta_{t}-\eta _{x}^{2},\label{V}
\end{equation}%
and
\begin{equation}
g_{2i+1}(t)=\frac{G_{2i+1}\zeta _{x}^{2}}{\rho
^{2i}},\;\;\;\;i=1,2,3,... \label{non}
\end{equation}

We introduce $a(t)=\sqrt{\tau _{t}}$ to get from \eqref{a} the relation 
$\zeta =a(t)x+b(t)$. This is important because through $\zeta$ we can now determine the
width of the localized solution in the form $1/a(t)$ and its center-of-mass position as $-b(t)/a(t)$;
see, e.g., Ref.~{\cite{BB}}. In the same way, through the use of (\ref{b}) one can write
\begin{equation}
\eta (x,t)=-\frac{a_{t}}{2a}x^{2}-\frac{b_{t}}{a}x+c(t).  \label{eta}
\end{equation}%
Thus, using (\ref{eta}) in (\ref{c}) one gets $\rho =\sqrt{a}$. Next, we 
rewrite the potential in the form
\begin{equation}
V(x,t)=f_{1}(t)x^{2}+f_{2}(t)x+f_{3}(t), \label{pot}
\end{equation}
with
\bes
\begin{eqnarray}
f_{1}(t)=\frac{a_{tt}}{2a}-\frac{a_{t}^{2}}{a^{2}},  \label{f1}
\end{eqnarray}%
\begin{eqnarray}
f_{2}(t)=\frac{b_{tt}}{a}-\frac{2a_{t}b_{t}}{a^{2}},  \label{f2}
\end{eqnarray}%
\begin{eqnarray}
f_{3}(t)=-c_{t}-\frac{b_{t}^{2}}{2a^{2}}.  \label{f3}
\end{eqnarray}%
\ees
In this case, $g_{2i+1}(t)=G_{2i+1}a^{2-i}$, for $i=1,2,3,...$, 
and the nonlinearities assume a simple form. However, we note 
that in the above general procedure one needs that $g_5=G_5$, 
suggesting that the strength of the quintic nonlinearity of 
the GPE \eqref{GPE} has to be constant. 

Similar ideas have been used recently in Refs.~\cite{SH,SH2,SHB,BB,BBC,ABC}. There one changes the nonautonomous 
NLSE to an autonomous NLSE depending on a single coordinate, in the form of a {\it static-like} ordinary differential
equation supporting soliton solutions. Such soliton solutions can be modulated via spatial and temporal dependences
associated with the potential and nonlinearity present in the nonautonomous NLSE. In this way, the soliton
can be modulated in a time-oscillating pattern, as a breather solution; see, e.g., Ref~{\cite{BB}} for a recent
investigation. In our procedure, however, we maintain the nonlinear equation in the form given by Eq.~(\ref{NLSE}),
which is a partial differential equation which possesses breather solution even in the absence of the spatial
and temporal modulations of the potential and cubic nonlinearity. This introduces an interesting difference because
now we are directly controlling the breather solutions. We explore this possibility in the next section,
investigating several distinct models.

%%%%%%%%%%%%%%%%%%%%%%%%%%%%%%%%%
\section{Specific models}
%%%%%%%%%%%%%%%%%%%%%%%%%%%%%%%%%
Let us now turn attention to the case of cubic nonlinearity, 
with the strength $g_3(t)=g(t)$ and no other nonlinearity,
which occurs for BECs with strong two-body scattering such as Rb or Na by use of 
Feshbach-resonance. In this case, the integrability of the NLSE (\ref{NLSE}) for 
cubic nonlinearity allows that one obtains breather solutions with distinct profile. 
A specific form of the two-soliton breather solution is obtained for $G_3=G=-1$, 
which corresponds to the explicit solution \cite{SatsumaPTPS74}
\begin{equation}
\Phi(\zeta,\tau )=\frac{4(\cosh(3\zeta )+3e^{4i\tau }\cosh (\zeta))e^{i\tau /2}}
{(\cosh (4\zeta )+4\cosh (2\zeta )+3\cos (4\tau ))}.
\label{sol1}
\end{equation}
It presents the simple form $\Phi(\zeta,0)=2\,{\rm sech}(\zeta)$ at $\tau=0$.
This is the simplest case, which we use to investigate distinct models below.

%%%%%%%%%%%%%%%%%%%%%%%
\begin{figure}[t]
\includegraphics[width=4.2cm]{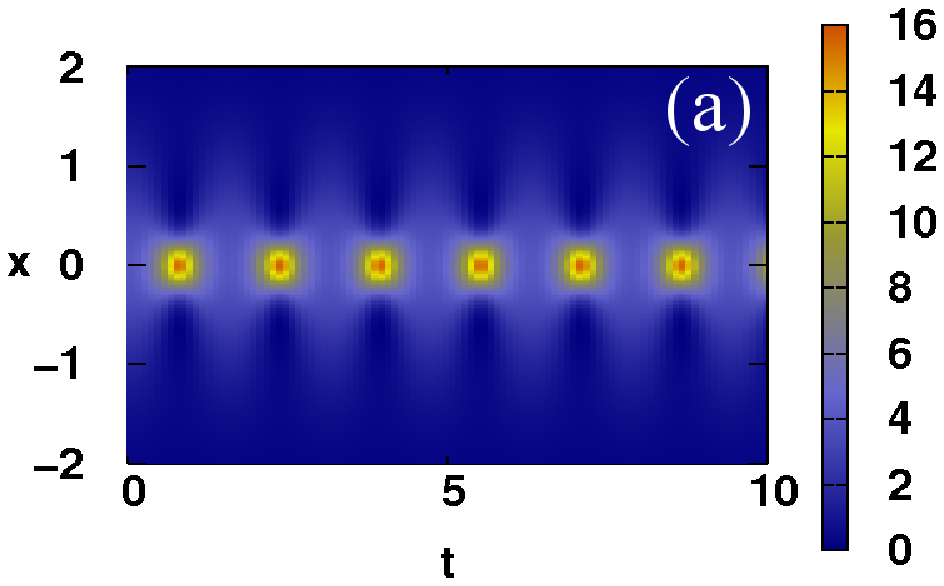}
\includegraphics[width=4.2cm]{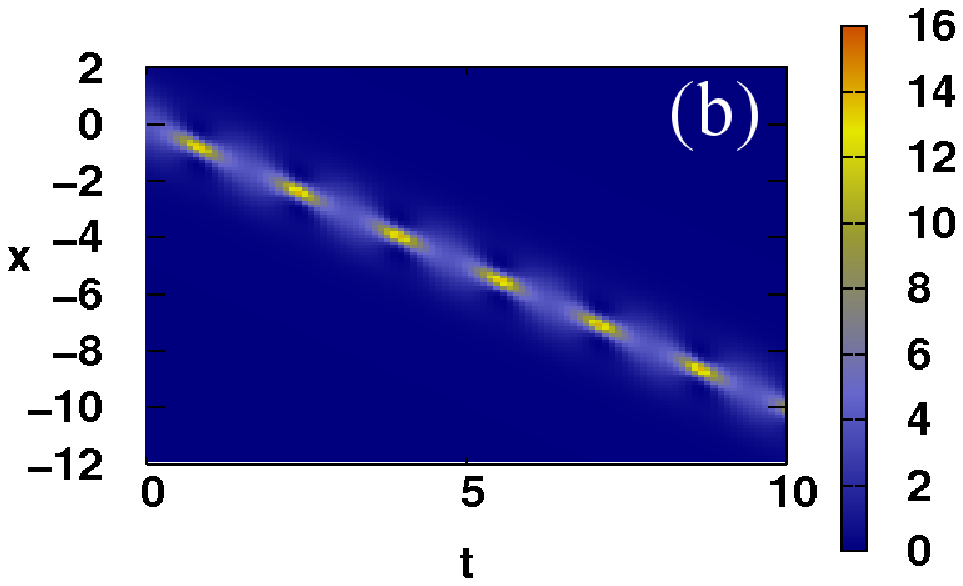}
\caption{(Color online) Breather solutions of the GPE \eqref{GPE}, in 
the case of a vanishing potential and constant cubic strength. In (a) we display 
the case $a\!=\!1$ and $b\!=\!0$ and in (b) $a\!=\!1$ and $b\!=\!t$, making 
the breather to move in space.}
\label{F1}
\end{figure}
%%%%%%%%%%%%%%%%%%%%

%%%%%%%%%%%%%%%%%%%%%%%%%%%%%%%%%%%%%%%%%
\begin{figure}
\includegraphics[width=6.0cm]{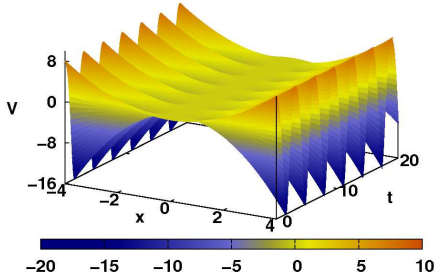}
\caption{(Color online) Plot of the potential shown in \eqref{p2}, using the 
parameters $a=1/(1+\cos^2(t))$ and $b=0$.}
\label{F2}
\end{figure}
%%%%%%%%%%%%%%%%%%%%%%%%%%%%%%%%%%%%%%%%%%

%%%%%%%%%%%%%%%%%%%%%%%%%%%%%%%%%%%%%%%
\subsection{Vanishing potential}

Firstly, we study the free evolution with vanishing potential 
and constant cubic nonlinearity strength. This case gives the standard breather
solution, with the NLSE with zero potential and constant coefficients. Thus, 
we consider $a=1$, which corresponds to $g=-1$ and $f_{1}=0$, and we take
$c_{t}=-b_{t}^{2}/2a^{2}$, leading to $f_{3}=0$. We can make $f_{2}=0$ for 
$b=0$ and for $b=t$. These breathers are standard solutions and they are
plotted in Figs. \ref{F1}a and \ref{F1}b, respectively. We note from 
\eqref{F1} that for $b=0$, the breather is localized in space and oscillates 
in time; however, for $b=t$ the corresponding center of mass moves in 
space with oscillating time behavior. This behavior is related to Eq.~(\ref{An}),
and to the nonzero phase in Eq.~(\ref{eta}), together with the fact that here $\rho$
is a constant and the form of the solution $\Phi$ in Eq.~(\ref{sol1}) also contributes
with a nontrivial phase: the overall effect is that $b=t$ makes the condensate to move
away from its starting position, as it is explicitly shown in Fig.~\ref{F1}b. The
oscillation period and frequency of the breather is of approximately $1.57$ and $0.64$,
respectively. We obtain the limits for minimum and maximum amplitude equal to $4$ and $16$,
respectively, and the time width at half height is approximately $0.39$. Recall that we are using
dimensionless coordinates.

\subsection{Flying bird potential}

\begin{figure}
\includegraphics[width=4.2cm]{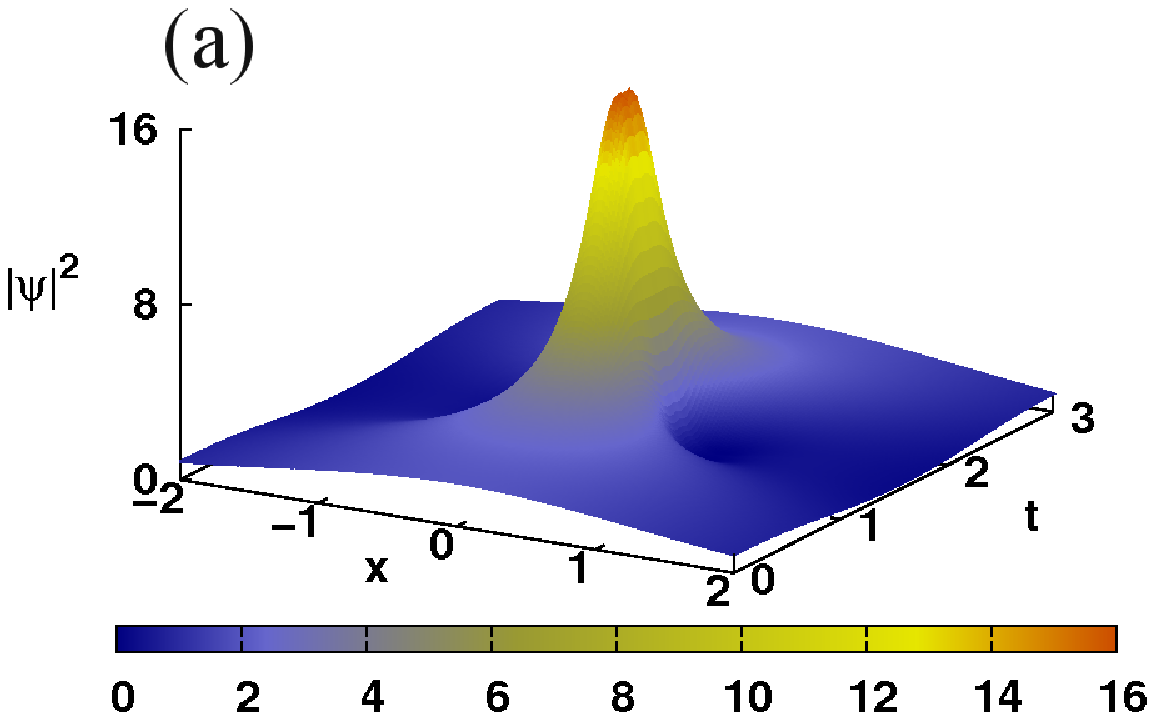}
\includegraphics[width=4.2cm]{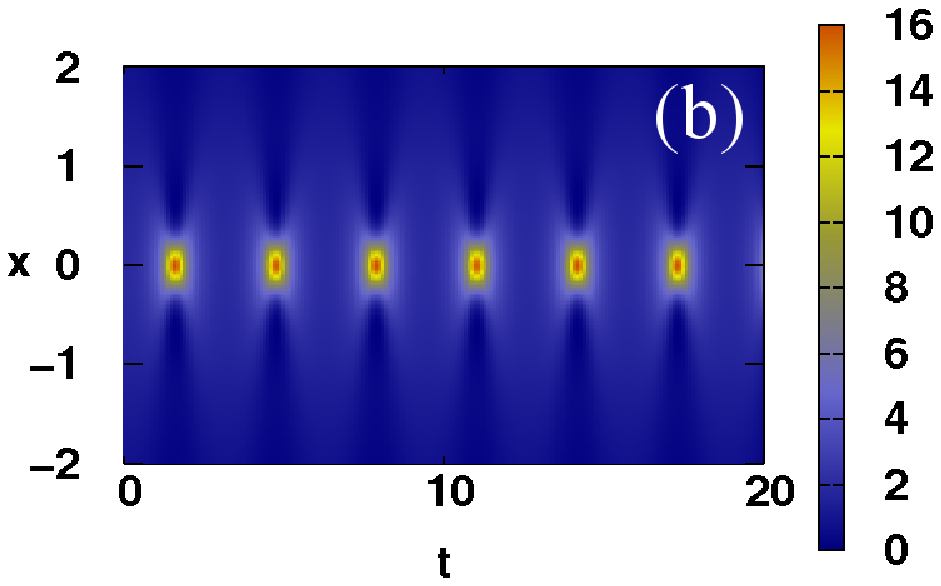}
\caption{(Color online) Plots of the (a) breather solution and (b) its 
profile, for the potential depicted in Fig.~{\ref{F2}}.}
\label{F3}
\end{figure}

Next, we deal with the specific case of a potential 
with space and time dependence, and a cubic nonlinearity with time-dependent 
strength. Here we choose the nonlinearity in the form $g=-1/(1+\cos ^{2}(t))$. 
To get a simple potential we take $b=0$ and $c=0$, which make $f_{2}=0$ and $f_{3}=0$.
In this case we get to the potential
\be
V=\frac{2\cos(2t)}{3+\cos(2t)}\,x^{2}.\label{p2}
\ee
Figs.~\ref{F2} and \ref{F3} show the potential and the breather solution, 
$\left\vert\psi\right\vert^{2}$, respectively. This potential changes from 
attractive to expulsive behavior controlled periodically \cite{Exp1,Exp2,Exp3}. 
We call this the flying bird potential, since it simulates the wings of a 
flying bird. Here we note that, differently from the first case, the periodic 
modulation does affect the periodic behavior of the breather solution. 
This interesting behavior appears from Figs.~\ref{F1} and \ref{F3}: there 
one sees that the breather exhibits a reduction in its oscillation frequency, 
which is modulated through the trapping potential and the strength of the 
cubic nonlinearity. 

We see that the oscillation period and frequency 
of the breather is of approximately $3.14$ and $0.32$, respectively. Note 
that the breather frequency was reduced by half when compared with the 
former case. Also, we obtain the limits for minimum and maximum amplitude
equal to $2$ and $16$, respectively, and the time width at half height is
approximately $0.54$.

\begin{figure}
\includegraphics[width=4.2cm]{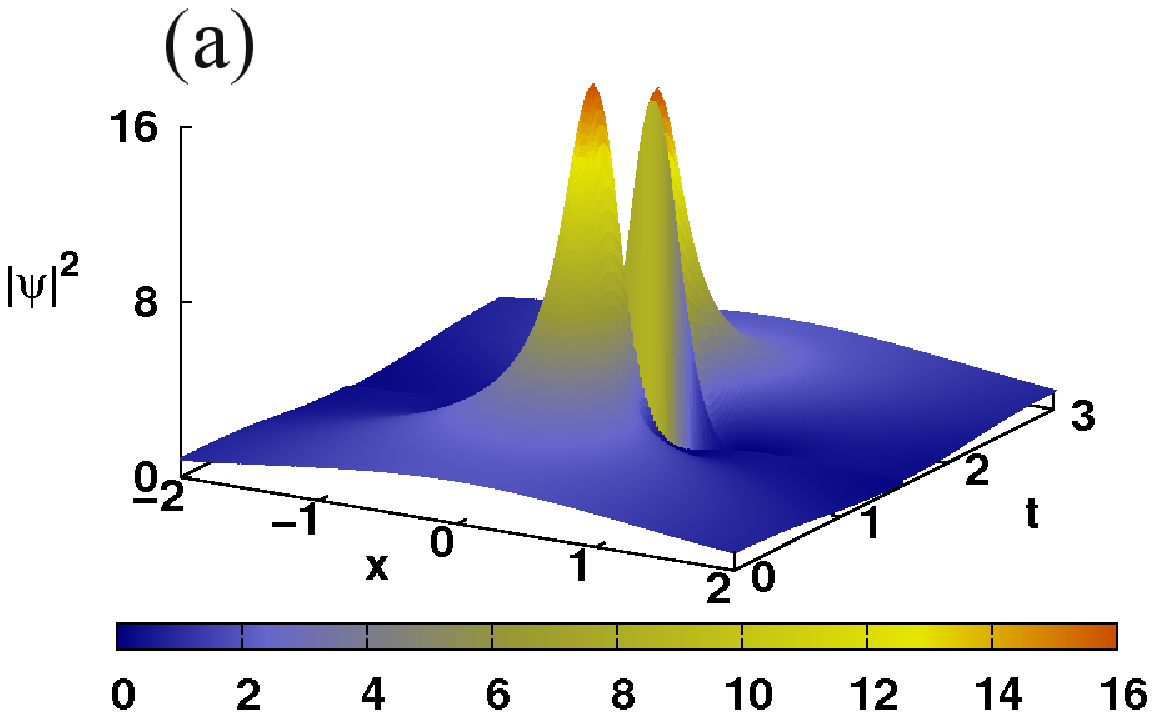}\includegraphics[width=4.2cm]{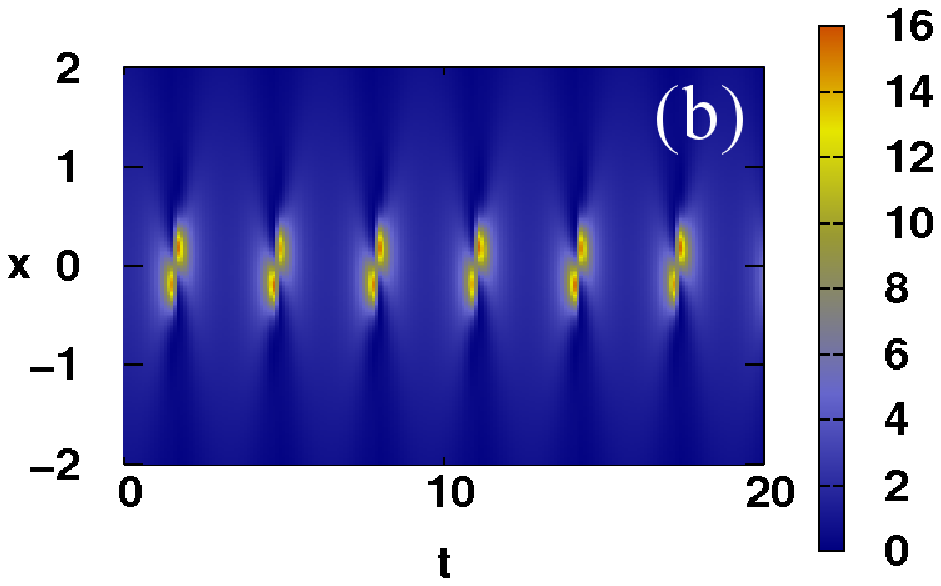}
\caption{(Color online) Plots of the (a) breather solution and (b) its profile. 
Here we are using $a$ and $b$ as in Fig.~{\ref{F2}}.}
\label{F4}
\end{figure}

The procedure can induce another important behavior to the solution. 
We can make the breather to move with the choice of $b$ such that
$b=-\tan(t)/16(\tan^{2}(t)+2)+3\sqrt{2}\arctan(\tan(t)/\sqrt{2})/32$, 
to get $f_{2}=0$; also, the choice $c_{t}=-b_{t}^{2}/2a^{2}$ makes $f_{3}=0$. 
In this way, using the same nonlinearity above the potential is also the same 
shown in (\ref{p2}). But now the breather present a nontrivial phase in (\ref{eta}),
and it presents an interesting behavior: it seems to split in two parts in the time coordinate.
This is different from the spatial splitting which was studied before in \cite{S04,Y09}. Our solution
is depicted in Fig.~\ref{F4}, and the breather seems to have two distinct time scales.

%%%%%%%%%%%%%%%%%%%%%%%%%%%%%%%%%%%%%%%%%%%%%%%
%%%%%%%%%%%%%%%%%%%%%%%%%%%%%%%%%%%%%%%%%%%%%%%
%%%%%%%%%%%%%%%%%%%%%%%%%%%%%%%%%%%%%%%%
%%%%%%%%%%%%%%%%%%%%%%%%%%%%%%%%%%%%

%%%%%%%%%%%%%%%%%%%%%%%%%
\begin{figure}
\includegraphics[width=4.2cm]{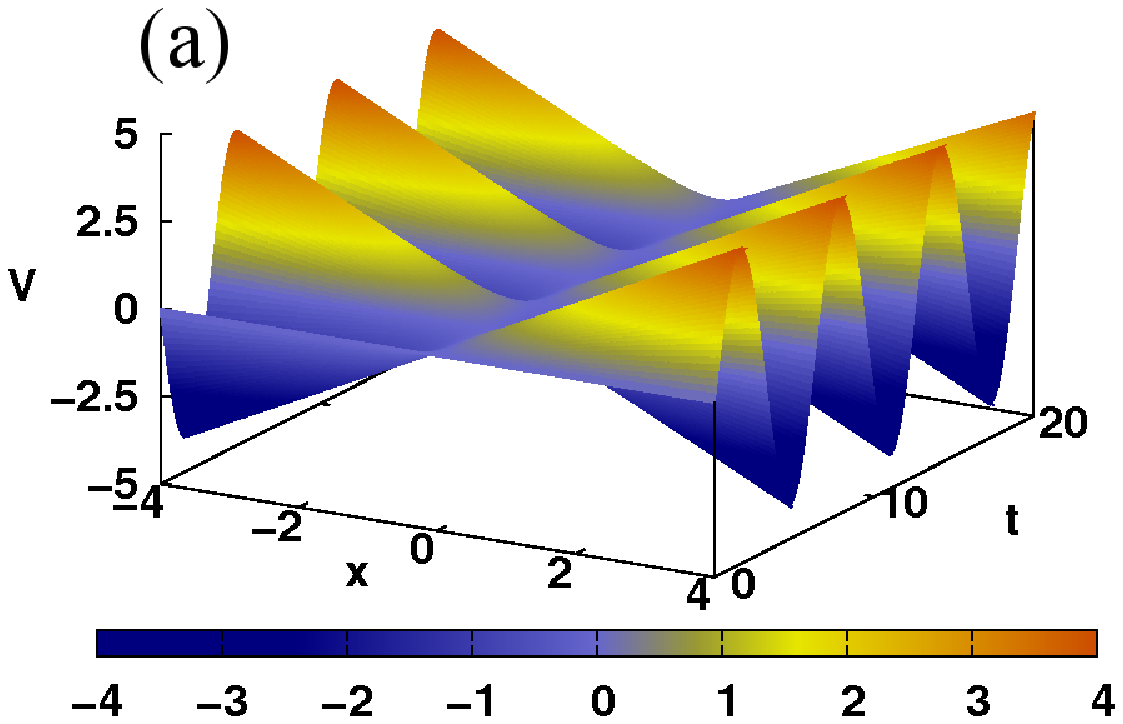}\includegraphics[width=4.2cm]{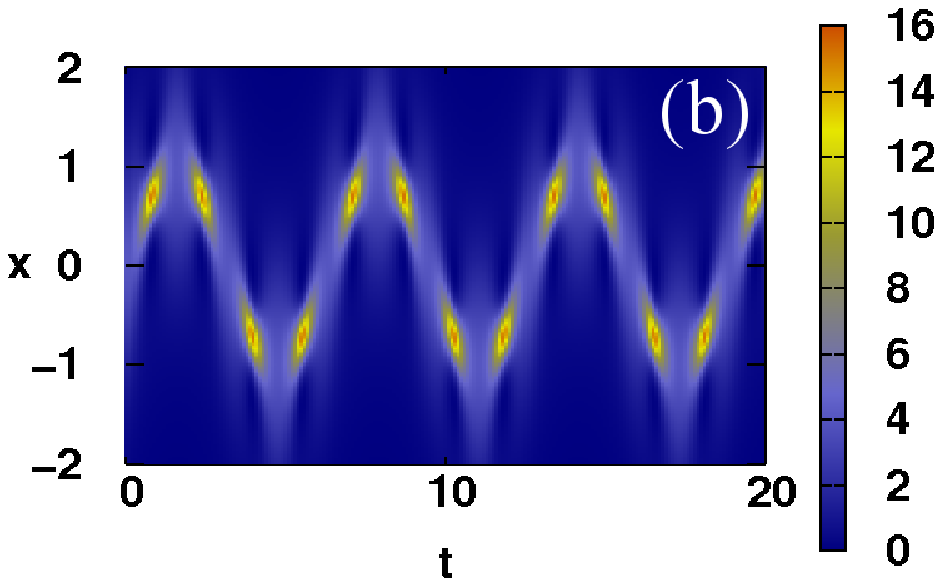}
\caption{(Color online) Plots of the (a) potential shown in \eqref{p3} and (b) 
the corresponding moving breather solution for $a$ constant and $b=\sin(t)$.}
\label{F5}
\end{figure}
%%%%%%%%%%%%

%%%%%%%%%%%%%%%%%%%%%%%%%%%%%%%
\subsection{Seesaw potential}

Another example is to verify the behavior of the moving breather 
solution in presence of a potential with linear spatial dependence. To this end 
we choose $f_{1}=0$, through the use of constant nonlinearity $a=1$, with $f_{3}=0$ 
for $c_{t}=-b_{t}^{2}/2a^{2}$. Thus, $b=-\sin (t)$ corresponds to the potential
\begin{equation}
V=\sin (t)\,x.\label{p3}
\end{equation}
This potential has a seesaw behavior, as it is displayed in Fig.~\ref{F5}a. 
In Fig.~\ref{F5}b we show the zig-zag pattern of the corresponding breather 
solution. In this case, the period of oscillation and frequency of the breather 
solution are $1.57$ and $0.64$, respectively; note that these values are the 
same of the first case. The amplitude now oscillates between $4$ and $16$ and the 
time width at half height is approximately $0.41$.

%%%%%%%%%%%%%%%%%%%%%%%%%%%%%%%%%%%%%%
\subsection{Another potential}

\begin{figure}
\includegraphics[width=6.0cm]{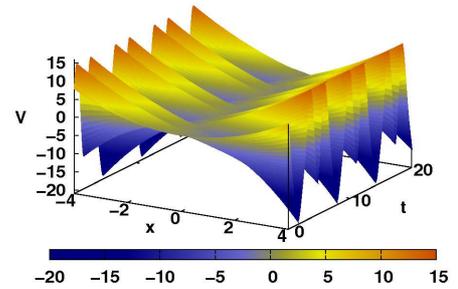}
\caption{(Color online) Plot of the potential shown in \eqref{pot15}.}
\label{F6}
\end{figure}

\begin{figure}
\includegraphics[width=4.2cm]{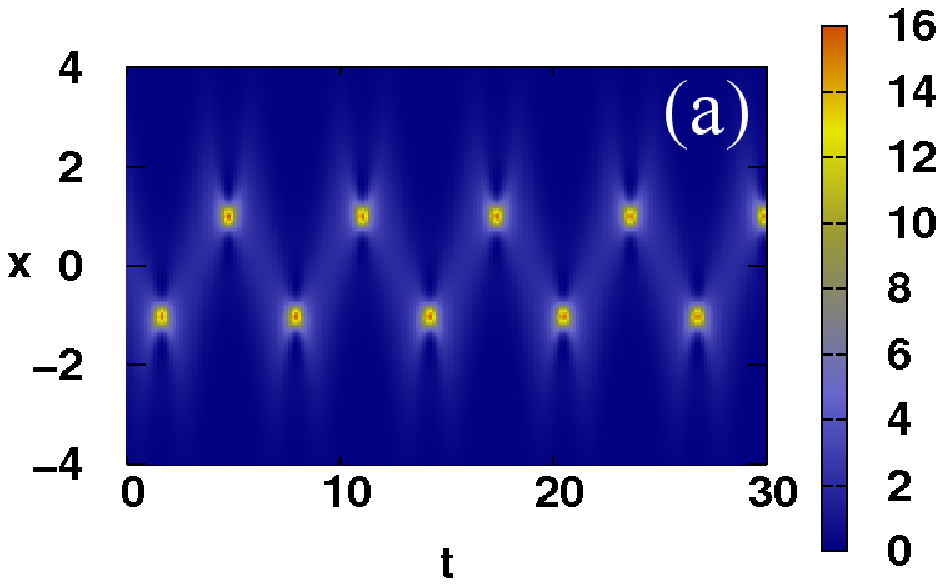}\includegraphics[width=4.2cm]{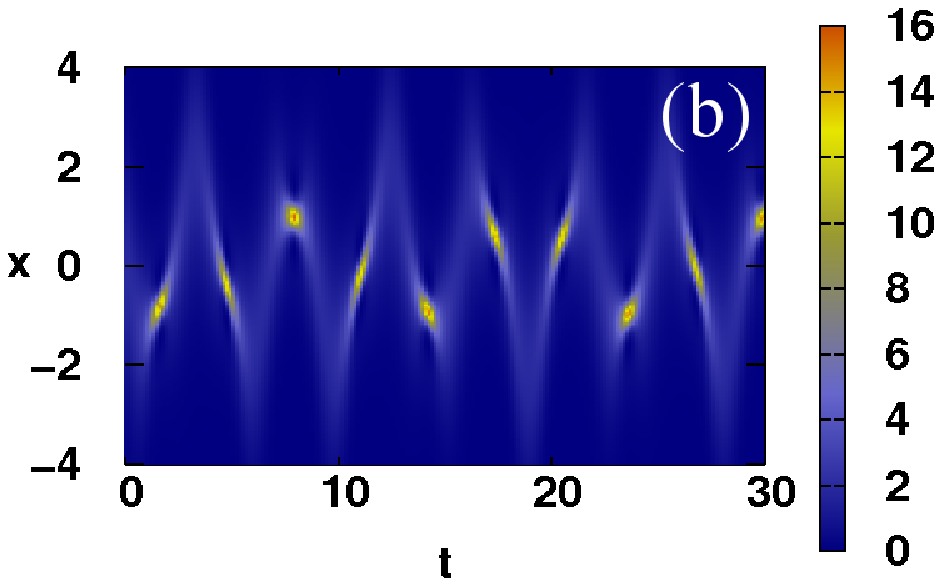}
\caption{(Color online) Plots of the profile of the moving breather solution,
showing the case of (a) periodic [$b=\sin(t)$] and (b) quasiperiodic [$b=\sin(\sqrt{2}t)$] patterns.} \label{F7}
\end{figure}

Other potentials can be studied for different values of $a$ and $b$. 
For instance, using the cubic nonlinearity strength $g=-1/(1+\cos ^{2}(t)),$ 
with $b=t$, and making $f_{3}=0$ we obtain the following potential 
\begin{equation}\label{pot15}
V\!=\!\frac{2\cos(2t)}{3+\cos(2t)}\,x^{2}-[1+5\cos^2(t)]\sin(t)\,x.
\end{equation}
It is displayed in Fig.~\ref{F6}. In this case, the breather solution presents 
an interesting moving pattern, as shown in Fig.~\ref{F7}a, for $b=\sin(t)$. 
Also, in Fig.~\ref{F7}b we display the behavior of the moving breather 
considering $b=\sin(\sqrt{2}t)$. With these choices, the breather can present 
periodic or quasiperiodic pattern, respectively, as it appears explicitly in 
the Figs.~\ref{F7}a and \ref{F7}b. In the case of periodic pattern,
we obtain the oscillation period of $3.14$ and frequency of $0.32$,
with the time width at half height given by $0.56$.

%%%%%%%%%%%%%%%%%%%%
%%%%%%%%%%%%%%%%%%%%%%%%%%%%
%%%%%%%%%%%%%%%%%%%%%%%%%%%%
%%%%%%%%%%%%%%%%%%%%

%%%%%%%%%%%%%%%%%%%%%%%%%%%%
\section{Numerical simulations}
%%%%%%%%%%%%%%%%%%%%%%%%%%%%

Starting with the GPE \eqref{GPE} in the 
case of cubic nonlinearity, we employ a numerical study focusing on the 
propagation of the modulated breather solution in the form given by Eq.~(\ref{An}). 
The numerical method is based on the split-step finite difference algorithm 
with a Crank-Nicolson method to solve 
the dispersive term in (\ref{GPE}). It is known that the Crank-Nicolson method 
is unconditionally stable \cite{Vesely94}.
In this way, using a short length of time one can obtain the true pulse propagation. 
Here, we used the time and space step-size of 
$0.0001$ and $0.01$, respectively. Thus, with a random perturbation of $5\%$ in the 
input profile the solution can 
expose its stable or unstable behavior. All the solutions were tested and appeared 
to be stable under the suggested random perturbations.
We analyzed the solutions for a very long time, until the rescaled time $t=10000$. 
The results strongly suggest experimental verification
of the above breather excitations in condensates.

Another important question to be considered is related to the validity of the one-dimensional
approximation to BECs, which we are using in this work. Evidently, we are assuming the regime
$a_s\ll a_{\bot}\ll\lambda\ll a_{||}$, where the Eq.~(\ref{GPE}) is supposed to be valid. However,
a more general possibility appears when one thinks of considering the quasi-one-dimensional dynamics,
which can be treated through a nonpolynomial nonlinear equation, as shown, for instance, in the recent
works \cite{NP1,NP2,NP3,NP4,NP5}. In this case, the \textit{Ansatz} used in Eq.~(\ref{An}) does not
transform the nonautonomous equation into an autonomous one.

\begin{figure}
\includegraphics[width=7cm]{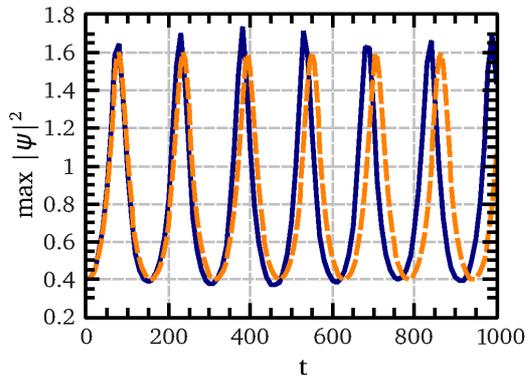}
\caption{(Color online) Plots of max $|\psi|^2$ for comparison between the cubic (orange, dashed line)
and nonpolynomial (blue, solid line) nonlinearities, considering $g=-0.1$ and $V=0$.}
\label{F8}
\end{figure}

To get further information on this issue, let us compare the results obtained in the present study with a numerical
investigation of the quasi-one-dimensional equation in presence of nonpolynomial nonlinearity given by \cite{NP1,NP2,NP3,NP4,NP5}
\begin{equation}
i\psi_t=-\frac{1}{2}\psi_{xx}+V(x,t)\psi+
\frac{1+(3/2)g(x,t)|\psi|^2}{\sqrt{1+g(x,t)|\psi|^2}}\psi.
\label{np}
\end{equation}
Note that for a weak-coupling regime, i.e., $g(x,t)|\psi|^2\ll1$, one can easily obtain the effective 
one-dimensional Eq.~(\ref{GPE}); see e.g., \cite{NP4} for more details on this issue. In our model,
this weak-coupling regime is contemplated when one decreases $a(t)$, i.e., when one increases the width
of the localized solution. As a simple example, let us compare the case {\bf A} of a vanishing potential,
studied above, with that obtained numerically through Eq.~(\ref{np}). We use the solution (\ref{sol1}) at $t=0$
as initial input in (\ref{np}) with $a=0.1$ and $b=0$, which correspond to $g=-0.1$ and $V=0$. The results are
shown in Fig. \ref{F8}. Note that the solid (blue) curve presents a slight delay, compared to the dashed (orange) curve
which represents case {\bf A}, of vanishing potential. We believe that this delay is due to the ${\rm max}\,(g|\psi|^2)\simeq0.16$,
which should be interpreted as $\ll1$. The results depicted in Fig.~\ref{F8} in solid (blue) line and in dashed (orange) line
reinforces the present study, showing that the one-dimensional approximation used in this work is reliable.

%%%%%%%%%%%%%%%%%%%%%%%%%%%%%%%%%%%%%%%%%%%
%%%%%%%%%%%%%%%%%%%%%%%%%%%%%%%%%%%%%%%%%%%%%%%%%%%%%%

%%%%%%%%%%%%%%%%%%%%%%%%%%
\section{Ending comments}
%%%%%%%%%%%%%%%%%%%%%%%%%%

In summary, we have studied the modulation of breather excitations in a nonautonomous GPE. 
We have used an {\it Ansatz} to transform the GEP into an autonomous NLSE which supports breathers.
The procedure allowed us to investigate a diversity of solutions, each one with its distinctive behavior.
The standard breather behavior appears in the simple case, with vanishing potential and constant
nonlinearity strength. But we could find other features, such as the moving, and the periodic
or quasiperiodic behavior. 

We have studied stability numerically, with the results showing that all the breather excitations
which we have found are stable against random perturbation  on the input profile of the solutions.
Also, we have investigated a simple case of nonpolynomial potential, and the results show that the
one-dimensional approximation which we have used in the work is reliable. The present study opens several
new possibilities, in particular on the experimental search for breather excitations in cigar-shaped
condensates, and on extensions of the procedure to condensates with more general nonlinearities and in 
higher dimensions.

\section*{Acknowledgments}

The authors would like to thank CAPES, CNPq, and FUNAPE/GO for partial financial support.

%%%%%%%%%%%%%%%%%%%%%%%%%%%%%%%%%%%%%%%%%%%%%%%%%%%%%%%%%

\end{document}